\documentclass[twocolumn,superscriptaddress,showpacs,prb]{revtex4-1}    

\usepackage{verbatim}
\usepackage[utf8]{inputenc}
\usepackage{amsmath,amssymb,amsfonts}
\usepackage{color}
\usepackage{hyperref}
\usepackage{filecontents}
\usepackage{graphicx}
\usepackage{soul}
\usepackage{bm}
\usepackage{makeidx}
\usepackage{txfonts}
\usepackage{mathtools}
\makeindex

\newcommand{\U}{{\mathrm{U}}}
\renewcommand{\L}{{\mathrm{L}}}
\renewcommand{\d}{{\mathrm{d}}}
\renewcommand{\H}{\mathcal{H}}

\begin{document}
\title{Transport through a quantum spin Hall antidot as a spectroscopic probe of spin textures}
\author{Alexia Rod}
\affiliation{Institute for Theoretical Physics, Technische Universit\"at Dresden, 01062 Dresden, Germany}
\affiliation{Physics and Materials Science Research Unit, University of Luxembourg, L-1511 Luxembourg}
\author{Giacomo Dolcetto}
\affiliation{Physics and Materials Science Research Unit, University of Luxembourg, L-1511 Luxembourg}
\author{Stephan Rachel}
\affiliation{Institute for Theoretical Physics, Technische Universit\"at Dresden, 01062 Dresden, Germany}
\author{Thomas L. Schmidt}
\affiliation{Physics and Materials Science Research Unit, University of Luxembourg, L-1511 Luxembourg}


\begin{abstract}
We investigate electron transport through an antidot embedded in a narrow strip of two-dimensional topological insulator. We focus on the most generic and experimentally relevant case with broken axial spin symmetry. Spin-non-conservation allows additional scattering processes which change the transport properties profoundly. We start from an analytical model for noninteracting transport, which we also compare with a numerical tight-binding simulation. We then extend this model by including Coulomb repulsion on the antidot, and we study the transport in the Coulomb-blockade limit. We investigate sequential tunneling and cotunneling regimes, and we find that the current-voltage characteristic allows a spectroscopic measurement of the edge-state spin textures.
\end{abstract}


\maketitle

\section{Introduction}

Two-dimensional topological insulators (2D TIs) behave as band insulators in the bulk but host gapless 1D edge states.~\cite{hasan10,qi11RMP83_1057} Experimentally, 2D TIs and their edge states have been investigated mostly in HgTe/CdTe quantum wells, as well as in InAs/GaSb heterostructures\cite{koenig-07s766,roth09,knez11,knez14,nowack13,spanton14} and evidence for the expected ballistic edge transport and the quantum spin Hall (QSH) effect has been found. In contrast to ordinary one-dimensional spin-$\tfrac{1}{2}$ electron systems, such as quantum wires, the edge channels of 2D TIs consist of a single pair of counter-propagating electronic modes.~\cite{kane-05prl146802,wu06,xu06} Time-reversal symmetry then severely impedes backscattering in the edge states, rendering them robust to disorder and weak interactions.

The simplest models for 2D TIs predict 1D edge channels in which electrons with opposite spins propagate in opposite directions.\cite{bernevig-06s1757} As helicity (i.e., the projection of the electron's spin operator on its momentum) is then conserved on a given edge, such systems are called helical 1D systems. However, while time-reversal symmetry is expected to be essential for the protection of gapless helical edge states, spin conservation is not. A plethora of effects, such as Rashba spin-orbit coupling, bulk inversion asymmetry, or structural inversion asymmetry, give rise to effective edge-state Hamiltonians without conserved spin.\cite{kane-05prl226801,liu08,rothe-12njp065012,liu11PRL107_76802}

In the presence of spin-symmetry breaking, left- and right-moving eigenstates can be almost arbitrary linear combinations of spin-up and spin-down electrons.\cite{schmidt-12prl156402} Time-reversal symmetry merely ensures that counter-propagating eigenstates with the same energy have opposite spin orientations, but it makes no statement relating eigenstates with different energies. Hence, the most generic helical system can be thought of as a helical channel in which the spin quantization axis can rotate with momentum, which makes the spin texture of a 2D TI edge state nontrivial even in the presence of time-reversal symmetry.\cite{dolcetto13b,dolcetto14} Recently, this spin texture was calculated for a number of realized and proposed 2D topological insulators based on their effective Hamiltonians.\cite{rod15PRB91_245112} Such generic helical liquids are the most general 2D TI edge states, characterized by time-reversal but no additional symmetries.

A nontrivial spin texture leads to interesting effects. Firstly, while zero-energy observables are insensitive to the spin texture, scattering processes at finite energies are greatly affected by the existence of right-movers and left-movers with non-orthogonal spins. This gives rise, for instance, to increased backscattering and thus a deviation from the quantized edge channel conductance at finite temperatures.\cite{schmidt-12prl156402,kainaris14,dolcetto15review} Another consequence of spin-non-conservation is the appearance of novel umklapp scattering processes that can gap out the spectrum even in the presence of time-reversal symmetry.\cite{orth15} Moreover, the spin texture can in principle be tuned locally by the application of a perpendicular electric field.\cite{rothe-12njp065012} In that case, coupling edge states with different spin textures has been shown to lead to new transport effects.\cite{orth13,rod15PRB91_245112,vandyke16}

In this article, we will investigate QSH antidots, i.e., non-topological regions (such as holes) embedded in a narrow strip of a 2D TI. In this case, the helical edge states propagating around the antidot can be tunnel-coupled to the helical systems propagating along the sample edges. Such a setup has a long history in the context of quantum Hall systems.\cite{goldman95,goldman08,geller97,komijani15} When embedded in 2D TIs, an antidot can be a useful tool to generate spin-polarized currents,\cite{dolcetto13} and thus to find evidence for the helicity of their edge states, as well as to explore nonlinear spin thermoelectric effects\cite{hwang2014nonlinear,lopez14}, entanglement~\cite{dolcetto16} or Kondo physics.\cite{posske13,posske14,rizzo16} The presence of multiple antidot-induced bound states was also shown to affect the transport properties of helical edge states, by inducing quantum percolation in the QSH bar.\cite{rui2012quantum}  Moreover, due to the potentially small sizes of the antidots, and the strong confinement of the electrons to 1D channels along their circumference, the Coulomb charging energy may be large. This provides a promising platform for studying the interplay between spin-orbit coupling and electron-electron interactions. Depending on the TI material at hand, antidots can in principle be realized either by lithographical patterning of the sample or by appropriate electrical gating.

In contrast to previous publications, the focus of this article will be on antidot transport in 2D TIs with a nontrivial edge-state spin structure. Our motivation is twofold: on the one hand, 2D TIs realized in InAs/GaSb or HgTe/CdTe systems are expected to have a nontrivial spin structure as a consequence of effects such as broken structural inversion asymmetry. Its effect should therefore be taken into account for a realistic modeling of antidot transport. On the other hand, it remains a challenge to directly measure the spin texture of edge states. We will show that using antidot geometries in the Coulomb-blockade regime, a spectroscopic measurement of the edge-state spin texture is possible by means of standard transport measurements.

The structure of this article is as follows: In Sec.~\ref{sec:model}, we will introduce the general model for an antidot embedded in a topological insulator without axial spin symmetry and present the low-energy Hamiltonian describing transport in the system. In Sec.~\ref{sec:non-int}, we will study transport in the absence of interactions on the antidot. In particular, we will present numerical results that allow us to fix the parameters of the analytical model. In Sec.~\ref{sec:int}, we will take into account the charging energy of the antidot, and we will present transport calculations in the sequential tunneling and cotunneling regimes. We present our conclusions in Sec.~\ref{sec:conclusions}.

\section{Model} \label{sec:model}

\begin{figure}[t!]
\centering
\includegraphics[scale=0.47]{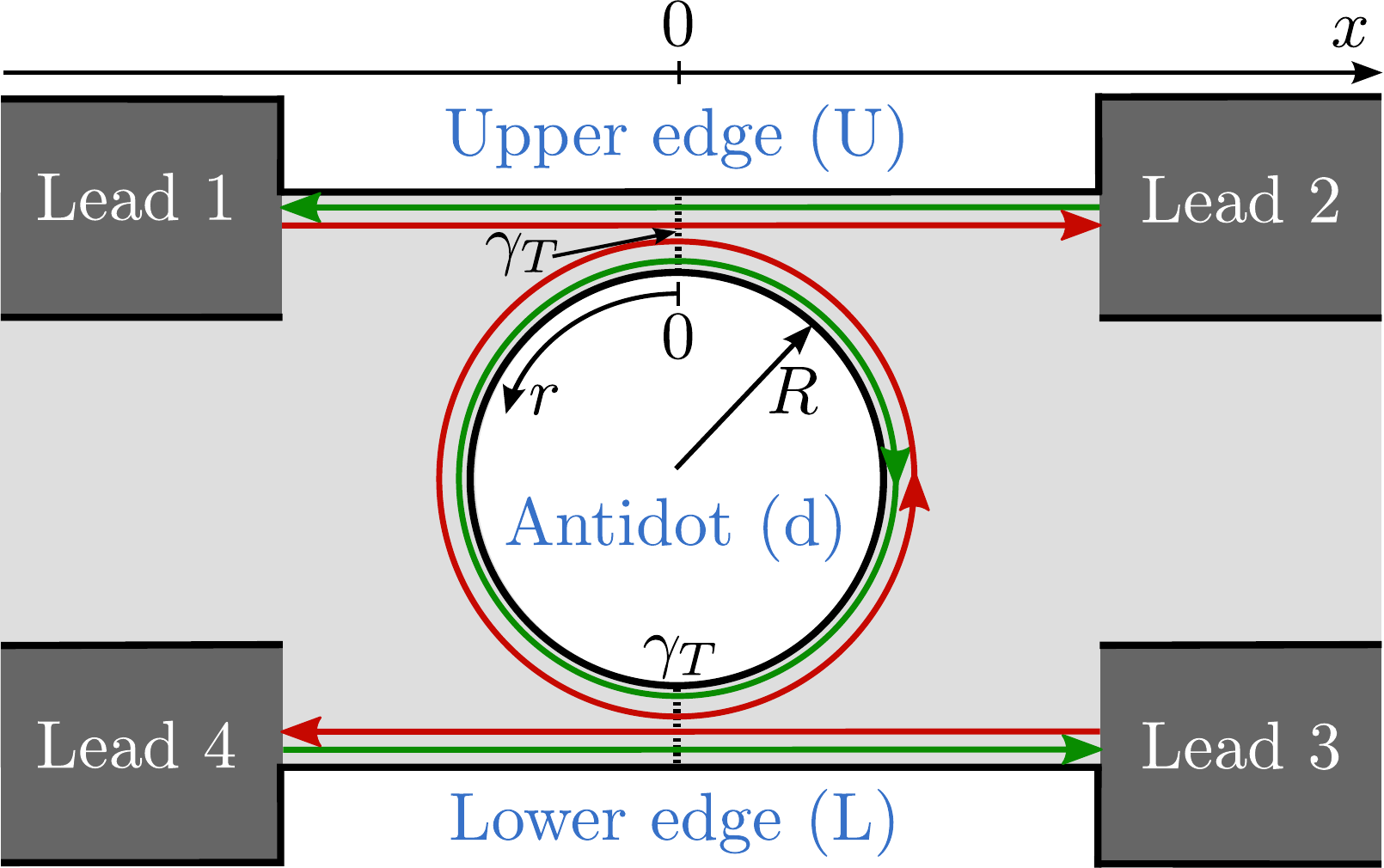}
\caption{ Sketch of the setup. The light gray area is the TI embedding an antidot of radius $R$. The dark gray areas are the leads. The tunneling processes (black dashed line) occur at $x=0$, $r=0$ and $x=0$, $r=\pi R$. The chirality of the edge states is indicated by the colors, '$+$' is red, '$-$' is green.}
\label{fig:setup}
\end{figure}

We consider an antidot geometry realized in a 2D TI, as schematically shown in Fig. \ref{fig:setup}.
This setup can be realized either by lithographically etching the sample or, in the case of an InAs/GaSb heterostructure, by gating the central portion of the bulk and thus bringing it to the trivial insulator regime. In both cases, a pair of helical edge states appear around the antidot ($\d$), in addition to the edge states present at the upper (U) and lower (L) edges of the QSH bar.
If the Fermi energy is tuned to lie within the bulk energy gap and the temperature is much lower than the gap itself, transport only occurs via the edge states, whereas the 2D bulk states are fully insulating.
The overlap between the edge-state wave functions gives rise to a finite tunneling probability between the edges and the antidot.

The total Hamiltonian in the presence of tunneling is
\begin{equation}\label{eq:Htot}
\H=\H_{\U}+\H_{\L}+\H_{\d}+\H_{\mathrm{d U}}+\H_{\mathrm{d L}},
\end{equation}
where $\H_{\mathrm{U(L)}}$ is the free Dirac Hamiltonian of the upper (lower) edge, $\H_{\d}$ is the Hamiltonian of the edge states around the antidot, and $\H_{\mathrm{d,U(L)}}$ is the tunneling Hamiltonian between the upper (lower) edge and the antidot.
Specifically, one has for the upper and lower edges,
\begin{equation}
\begin{split}
\H_{s}&=-is v_F\sum_{\alpha=\pm}\alpha\int_{-l/2}^{l/2}\!dx\, \Psi^\dagger_{s\alpha}(x)\partial_x \Psi_{s\alpha}(x),\label{hamiltonian-edge}
\end{split}
\end{equation}
where $s=\mathrm{U,L}\equiv +,-$. Moreover, $v_F$ is the Fermi velocity, and $l$ is the edge length. We mostly consider the limit $l\to\infty$, thus assuming the upper and lower edges to have a continuous spectrum, contrary to the antidot, whose energy levels are discrete with an energy separation $\approx v_F/R$, $R$ being the radius of the antidot.
In the presence of axial spin symmetry, the quantum number $\alpha$ corresponds to the spin polarization of the edge states, so that for instance spin-up electrons propagate to the right on the upper edge and to the left on the lower one (opposite for spin-down electrons).

Our aim is to investigate the more general experimental scenario in which axial spin symmetry is broken: in this case, the electron operators $\Psi^{\dagger}_{s\pm}(x)$ still correspond to chiral particles moving to the right (left) on the upper edge and to the left (right) on the lower one, but they are no longer eigenstates of the spin operator.
In momentum space it is possible to relate the chiral basis $c_{s\alpha k}= \int dx e^{-ikx}\Psi_{s\alpha}(x)/\sqrt{l}$ for $\alpha = \pm$ to the spin basis $c_{s\sigma k}= \int dx e^{-ikx}\Psi_{s\sigma}(x)/\sqrt{l}$ for $\sigma = \uparrow,\downarrow$ via a unitary transformation\cite{schmidt-12prl156402}
\begin{equation}
\begin{pmatrix}
c_{s+,k}\\
c_{s-,k}\\
\end{pmatrix}=B_{sk}^\dagger \begin{pmatrix}
c_{s\uparrow k}\\
c_{s\downarrow k}\\
\end{pmatrix}
.\end{equation}
The form of the momentum-dependent $\mathrm{SU}(2)$ matrix $B_{sk}$ is dictated by time-reversal symmetry and unitarity
\begin{equation}
B_{sk}=\begin{pmatrix}
\cos(\theta_{sk}) & -\sin(\theta_{sk})\\
\sin(\theta_{sk}) & \cos(\theta_{sk})
\end{pmatrix}
\label{Bk-matrix},
\end{equation}
where the function $\theta_{sk}$, which is even in $k$ because of time-reversal symmetry, measures the rotation of the spin-quantization axis~\cite{rod15PRB91_245112} on edge $s$ at momentum $k$. For realistic models and momenta near the Dirac point, it was shown\cite{rod15PRB91_245112} that one can usually use the approximation $\theta_{sk}\approx(k_s/k_{0s})^2$, where the parameter $k_{0s}$ represents the momentum scale over which the spin-quantization axis rotates and thus incorporates the information about the spin structure of the helical states.
Note that we allow in principle the upper and lower edge to have different spin structures with parameters $\theta_{\mathrm U k}$ and $\theta_{\mathrm L k}$.

The antidot Hamiltonian
\begin{equation}
\H_{\d}=-iv_F\sum_{\alpha=\pm}\alpha\int_0^{2\pi R}\!dr\, \Psi^\dagger_{\d\alpha}(r)\partial_r \Psi_{\d\alpha}(r)+E(n) \label{hamiltonian-antidot}
\end{equation}
is also characterized by a linear dispersion. However, due to its confinement, the charging energy contribution should be taken into account,
\begin{equation}
E(n)=\frac{E_{ c}}{2}\left(n-\frac{eV_g}{E_{c}}\right)^2,
\end{equation}
where $E_{c}$ is the Coulomb energy, $V_g$ is the gate voltage applied to the island, and $n=\sum_{\alpha}\int_0^{2\pi R}\!dr\, \Psi^\dagger_{\d\alpha}(r) \Psi_{\d\alpha}(r)$ is the number operator.
As for the edges, spin in general is not a good quantum number, so the operator $\Psi^\dagger_{\d\pm}(r)$ refers to electrons propagating clockwise/anticlockwise but without a well-defined spin polarization. Following what was done for the translationally invariant edges, we can define the most general $\mathrm{SU}(2)$ transformation in angular momentum space. It relates the chiral states $d_{\alpha j}=\int dre^{-ijr/R}\Psi^\dagger_{\d\alpha}(r)/\sqrt{2\pi R}$ for $\alpha = \pm$ to the spin-polarized ones $d_{\sigma j}=\int dre^{-ijr/R}\Psi^\dagger_{\d\sigma}(r)/\sqrt{2\pi R}$ for $\sigma = \uparrow,\downarrow$ as
 \begin{equation}
\begin{pmatrix}
d_{+,j}\\
d_{-,j}\\
\end{pmatrix}=\tilde{B}_{j}^\dagger \begin{pmatrix}
d_{\uparrow j}\\
d_{\downarrow j}\\
\end{pmatrix}
.\end{equation}
It is given by
\begin{equation}
\tilde{B}_{j}= \begin{pmatrix}
\cos(\theta_{j}) & -\sin(\theta_{j})\\
\sin(\theta_{j}) & \cos(\theta_{j})
\end{pmatrix}\label{Bj-matrix}
.\end{equation}

Tunneling to and from the antidot occurs near the coordinates $x=0$ and $r=0=:r_\U$ for the upper contact and at $x=0$ and $r=\pi R=:r_\L$ for the lower one (see Fig.~\ref{fig:setup}).
We start with the most general tunneling Hamiltonian containing both spin-preserving and spin-flipping terms, \cite{krueckl11}
\begin{align}
\H_{\d s} &= \sum_{\sigma\sigma'} \int dx dr \left[
\Psi^\dagger_{s\sigma}(x) \gamma^s_{\sigma\sigma'}(x,r) \Psi_{\d\sigma'}(r_s + s r) + \text{H.c.}\right]
\end{align}
where $s = \U,\L \equiv +,-$ and $\sigma,\sigma' \in \{\uparrow,\downarrow\}$. To limit the number of parameters, we assume the sample geometry to be symmetric about the $x$ axis, see Fig.~\ref{fig:setup}. Reflection symmetry about the $x$ axis is defined as $(x,y) \to (x,-y)$ and $(p_x, p_y) \to (p_x, -p_y)$. This entails the transformation rule $\sigma_z \to -\sigma_z$ for the spin quantum number. As a consequence, the field operators transform as $\Psi_{\U\sigma}(x)\rightarrow\Psi_{\L\bar{\sigma}}(x)$ and $\Psi_{\d\sigma}(r)\rightarrow\Psi_{\d\bar{\sigma}}(\pi R-r)$. Invariance of $\H_{\d \U} + \H_{\d \L}$ under this transformation leads to the four equations $\gamma^\U_{\sigma\sigma'} = \gamma^\L_{\bar{\sigma}\bar{\sigma}'}$, which allow us to eliminate $\gamma^\L$, and it leaves only the four functions $\gamma_{\sigma\sigma'} \equiv \gamma^\U_{\sigma\sigma'}$. In addition, we assume the tunnel Hamiltonian to respect time-reversal symmetry, which is local in space and acts on the edge states as $\Psi_{s\sigma}(x)\rightarrow\sigma \Psi_{s\bar{\sigma}}(x)$ for $s \in \{\U, \L, \d\}$ and $\sigma =\ \uparrow,\downarrow\ = +,-$. The tunnel Hamiltonian has time-reversal symmetry if $\gamma_{\uparrow\uparrow} = \gamma_{\downarrow\downarrow}$ and $\gamma_{\uparrow\downarrow} = - \gamma_{\downarrow\uparrow}$. This leaves us with two functions $\gamma_{\rm sc} \equiv \gamma_{\uparrow\uparrow}$ and $\gamma_{\rm sf} = \gamma_{\uparrow\downarrow}$ denoting the amplitudes of spin-conserving and spin-flip tunneling, respectively.\cite{dolcini11,dolcini15} The Hamiltonian now reads
\begin{align}
\H_{\d s} &= \sum_{\sigma} \int dx dr \big[
\Psi^\dagger_{s\sigma}(x) \gamma_{\rm sc}(x,r) \Psi_{\d\sigma}(r_s + s r) \notag \\
+&
s \sigma \Psi^\dagger_{s\sigma}(x) \gamma_{\rm sf}(x,r) \Psi_{\d\bar{\sigma}}(r_s + s r) +
\text{H.c.}\big].
\end{align}
We would like to point out that it is important to fix the form of the tunneling Hamiltonian by reflection symmetry and not inversion symmetry, the latter being defined as $(x,y) \to (-x, -y)$. Indeed, as we will show further below for the Kane-Mele model,\cite{kane-05prl146802,kane-05prl226801} in the presence of Rashba spin-orbit coupling, the bulk system remains invariant under reflection, whereas inversion symmetry is usually lost.

Next, we express the tunneling Hamiltonian in the basis of the chiral edge states. For this purpose, we Fourier-transform to momentum and angular momentum space and use the rotation matrices (\ref{Bk-matrix}) and (\ref{Bj-matrix}). Expressed in terms of the Fourier components of the tunneling amplitudes, we find
\begin{align}
\H_{\d s} =\ & \frac{1}{\sqrt{2 \pi R l}} \sum_{\sigma} \sum_{k,j} \sum_{\alpha\alpha'} \big[  \\
&
e^{i j r_s/R} \tilde{\gamma}_{\rm sc}(k,s j) (B^\dag_{s,k})^{\alpha\sigma} \tilde{B}_{j}^{\sigma \alpha'} c^\dagger_{s\alpha k}  d_{\alpha' j} \notag \\
+\ &
e^{i j r_s/R} \tilde{\gamma}_{\rm sf}(k,s j) (B^\dag_{s,k})^{\alpha\sigma} \tilde{B}_{j}^{\bar{\sigma} \alpha'} s \sigma c^\dagger_{s\alpha k}  d_{\alpha' j}+
\text{H.c.}\big].\notag
\end{align}
We can further simplify this by assuming that the Fourier components of the tunneling amplitudes $\tilde{\gamma}_{\rm{sc,sf}}(k,j)$ as well as the rotation matrices vary slowly as functions of $k$ and $j$. The former is justified if the tunneling happens locally on the scale of the Fermi wavelength. The latter assumption holds if temperature and applied bias voltage are small compared to $v_F k_{0\{\U,\L\}}$. In this case, we can replace these functions by their values at the Fermi energy and define
\begin{align}
    \gamma_T \cos(\theta_T) &= \tilde{\gamma}_{\rm sc}(k_F, j_F) \notag \\
    \gamma_T \sin(\theta_T) &= \tilde{\gamma}_{\rm sf}(k_F, j_F) \notag \\
    \tilde{B} &= \tilde{B}_{j_F} \notag \\
    B_{s} &= B_{s,k_F} \quad \text{for }s \in \{\U, \L\}
\end{align}
where $k_F = \mu/v_F$ and $j_F = \mu R/v_F$ are determined by the chemical potential $\mu$. The angle $\theta_T$ set the ratio between the tunneling amplitudes for spin-conserving and spin-flip tunneling, $\theta_T=\tan^{-1}( \tilde{\gamma}_{\rm sf}/ \tilde{\gamma}_{\rm sc})$. Then, we obtain by Fourier-transforming back to real space
\begin{equation}
\begin{split}
\H_{\d s}&=\gamma_T \sum_{\alpha \alpha^\prime}  \Psi^\dagger_{s \alpha}(0) \phi_{s\alpha\alpha'} \Psi_{\d \alpha^\prime}(r_s) + \mathrm{h.c.}, \\
\phi_{s\alpha\alpha'} &= \sum_{\sigma}(B_{s}^\dagger)^{\alpha\sigma}\left (\cos\theta_T\tilde{B}^{\sigma\alpha^\prime}+s \sigma \sin\theta_T\tilde{B}^{\bar{\sigma}\alpha^\prime}\right ).
\end{split}
\end{equation}
In the following we study how the spin structure of the helical edge states can be explored by means of transport properties. We begin by investigating the noninteracting case, which we can compare with numerical simulations on a lattice.

\section{Non-interacting antidot}\label{sec:non-int}

To investigate the transport properties in the absence of interactions we use the standard scattering matrix formalism.\cite{datta_book} After calculating the Heisenberg equations of motion $i\partial_t \Psi_{s\alpha} = [\Psi_{s\alpha},\H]$ and $i\partial_t \Psi_{\mathrm{d}\alpha} = [\Psi_{\mathrm{d}\alpha},\H]$ with respect to the Hamiltonian~\eqref{eq:Htot}, and by imposing plane-wave solutions for the states coming from (with amplitude $a_i$) and going to (with amplitude $b_i$) the contacts $i=1,\ldots,4$ (see Fig.~\ref{fig:setup}), we can find the scattering matrix relating $b_i=\sum_jS_{ij}a_j$ as
\begin{widetext}
\begin{equation}\label{eq:S}
S=\frac{1}{\left(1+\Gamma^2\right) \sin \phi +2 i \Gamma \cos \phi }\begin{pmatrix}
0 & \left(1-\Gamma^2\right) \sin\phi & 2i\Gamma\sin\Theta & -2i\Gamma\cos\Theta\\
\left(1-\Gamma^2\right) \sin\phi & 0 & -2i\Gamma\cos\Theta &-2i\Gamma\sin\Theta\\
-2i\Gamma\sin\Theta & -2i\Gamma\cos\Theta & 0 & \left(1-\Gamma^2\right) \sin\phi\\
-2i\Gamma\cos\Theta & 2i\Gamma\sin\Theta &  \left(1-\Gamma^2\right) \sin\phi &0
\end{pmatrix},
\end{equation}
\end{widetext}
with the dimensionless tunneling probability $\Gamma=|\gamma_T|^2/(4v^2_F)$.
The scattering matrix depends on the chemical potential through the phase factor $\phi=\pi R\mu/v_F$ and on the parameter $\Theta=\theta_{\mathrm{U}k_F}-\theta_{\mathrm{L}k_F}+2\theta_T$.
The current measured at the $i$-th contact can be evaluated from Eq.~\eqref{eq:S} using the Landauer-B\"{u}ttiker formula
\begin{equation}\label{eq:LB}
I_i=\frac{G_0}{e}\sum_j\int_{\mu+eV_j}^{\mu+eV_i} \mathclap{dE \, T_{ij}(E),}
\end{equation}
where $G_0=e^2/(2\pi)$ is the conductance quantum, $T_{ij}=\left\vert S_{ij}\right\vert^2$ are the elements of the transmission matrix, and $\{V_j\}$ are the bias potentials applied to the four contacts.
It is worth noting that the spin structure of the helical states on the antidot does not affect the transport properties. Nevertheless, the result does depend on the spin textures of the edges through $\Theta$. Therefore, if the latter can be tuned independently, for instance by applying an electric-field gradient, they can be directly resolved via a current measurement even in the noninteracting case. This result is analogous to what was found for a tunnel junction between two edges.\cite{orth13}
In contrast, in the homogeneous case, i.e., with the same spin structure $\theta_{\mathrm{U}k}=\theta_{\mathrm{L}k}$ on both edges, $\Theta=2\theta_T$ and the transmission matrix is uniquely determined by the ratio between spin-preserving and spin-flipping tunneling.
If the chemical potential coincides with an eigenenergy of the antidot $\mu= v_F j/R$ (a scenario which we will refer to as the resonant case), the phase factor $\sin \phi=0$ so that the incoming electron is fully transmitted across the antidot, while away from resonance one recovers the typical Lorentz-shaped transmission for transport through a quantum (anti)dot.

These results are confirmed by numerical transport simulations using the KWANT package.\cite{groth_kwant} To investigate the effects induced by breaking the axial spin symmetry we consider the Kane-Mele (KM) lattice model\cite{kane-05prl146802,kane-05prl226801} on the honeycomb lattice, which is defined as
\begin{eqnarray}\label{KM_Hamiltonian}
\H_{\rm KM}&=&-t\sum\limits_{\langle ij\rangle}c^\dagger_ic_j+i\lambda_{\rm SO}\sum\limits_{\langle\!\langle ij \rangle\!\rangle}\nu_{ij}c^\dagger_is^zc_j\nonumber\\
&&{}+i\lambda_{\rm R}\sum\limits_{\langle ij \rangle}c^\dagger_i(\boldsymbol{s}\times \hat{ \boldsymbol{d}}_{ij})_zc_j,
\end{eqnarray}
where $c_i=(c_{i\uparrow},c_{i\downarrow})$ is a two-component spinor, $\nu_{ij}=\pm 1$ is a factor that is $+1$ ($-1$) if the next-nearest-neighbor hopping from site $j$ to site $i$ corresponds to a right turn (left turn) in the honeycomb lattice, $\boldsymbol{s}$ is the spin operator, and $\hat{\boldsymbol{d}}_{ij}$ is the unit vector between the nearest-neighbor lattice sites $i$ and $j$. The parameters of the model are the hopping amplitude $t$, the intrinsic spin-orbit coupling $\lambda_{\rm SO}$, and the Rashba spin-orbit coupling (RSOC) $\lambda_{\rm R}$, which is responsible for the axial spin symmetry breaking. In the following we limit ourselves to uniform bulk parameters, which corresponds in the analytical model to the homogeneous case $\theta_{\mathrm{U}k}=\theta_{\mathrm{L}k}$.

The KM Hamiltonian with RSOC does not preserve inversion symmetry.\cite{fu07prb76_45302} Inversion does not affect the spin as the latter is a pseudovector. It does exchange the two sublattices forming the honeycomb lattice, but it leaves the phase $\nu_{ij}$ invariant. Hence, the kinetic and the intrinsic spin-orbit terms of the Hamiltonian are invariant. However, the RSOC term gets a minus sign under inversion, which destroys the inversion symmetry of the total Hamiltonian.

On the other hand, $\H_{\rm KM}$ has reflection symmetry. A reflection about the $x$ axis ($y \to -y$) will change the signs of the $x$ and $z$ components of the spin. Moreover, reflection symmetry swaps the sublattices. Hence, the kinetic term is invariant. For the spin-orbit part, the $z$ component of spin will pick up a minus sign, but $\nu_{ij}$ will change sign, too. For the RSOC term, the $y$ component ($x$ component) of the lattice vectors switches (does not switch) sign, but the $x$ component ($y$ component) of the spin also switches (does not switch) sign. Hence, a reflection about the $x$ axis leaves the Hamiltonian invariant.

\begin{figure}[t!]
\centering
\includegraphics[scale=0.6]{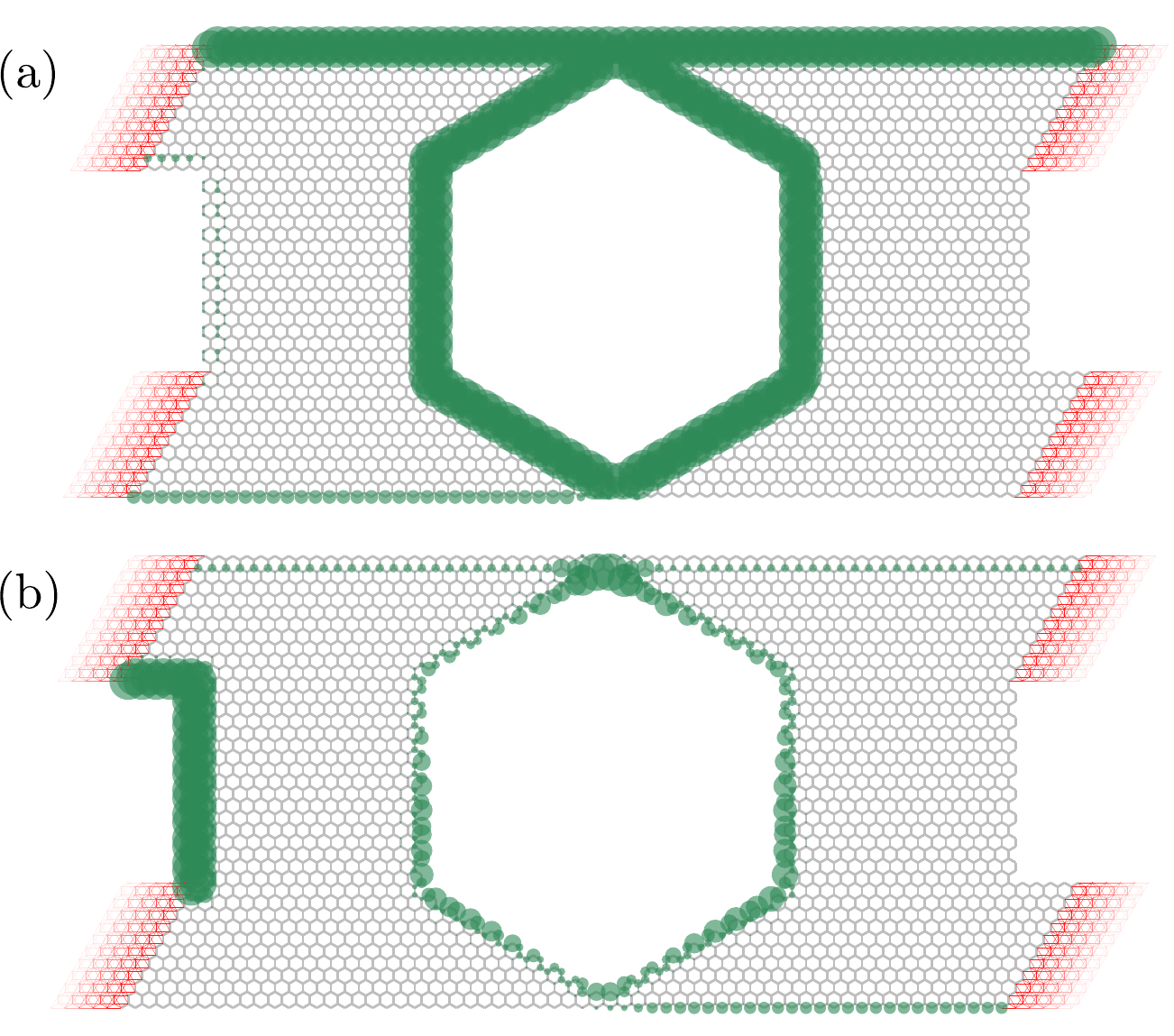}
\caption{Representative example of the electron densities (green dots) for scattering states originating from lead $1$. Panel (a) shows the up-spin propagation and panel (b) shows the down-spin propagation. The lattice is 60 lattice constants long and 34 lattice constants wide. The leads, shown in red, are 9 lattice constant wide. The model parameters are set to: $t=1$, $\lambda_{\rm SO}=0.2 t$, $\lambda_{\rm R}=0.1 t$ and $\mu=0.157t$.}
\label{fig:wavefunction}
\end{figure}

For the numerical simulation, we consider a finite lattice connected to four leads, as shown in Fig.~\ref{fig:wavefunction}. To avoid dangling bonds, we model the antidot as hexagonal-shaped. The numerical calculation provides access to the full scattering matrix, and without loss of generality we will discuss in the following the transmission of electrons injected from lead 1 to the other leads. The finite-length system provides, in addition to transport via the antidot, direct ballistic channels between the upper and the lower edge along the left and right edges of the sample. However, these are easy to distinguish from transport via the antidot. Fig.~\ref{fig:wavefunction} shows the weight of the spin-up and spin-down wavefunctions for small RSOC ($\theta_{sk_F}\ll 1$) on the antidot and the leads.

\begin{figure}
\centering
\includegraphics[width=\columnwidth]{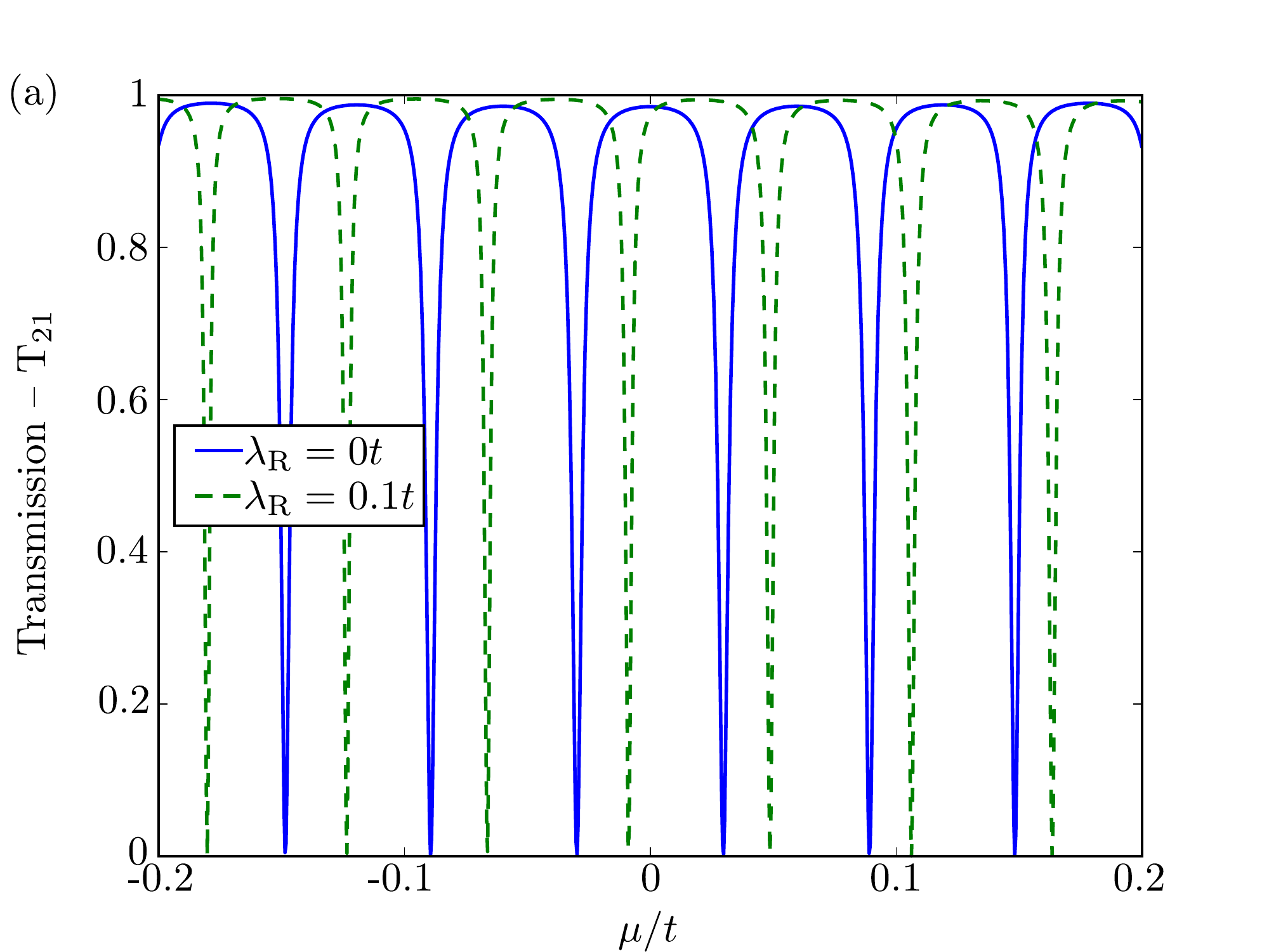}
\includegraphics[width=\columnwidth]{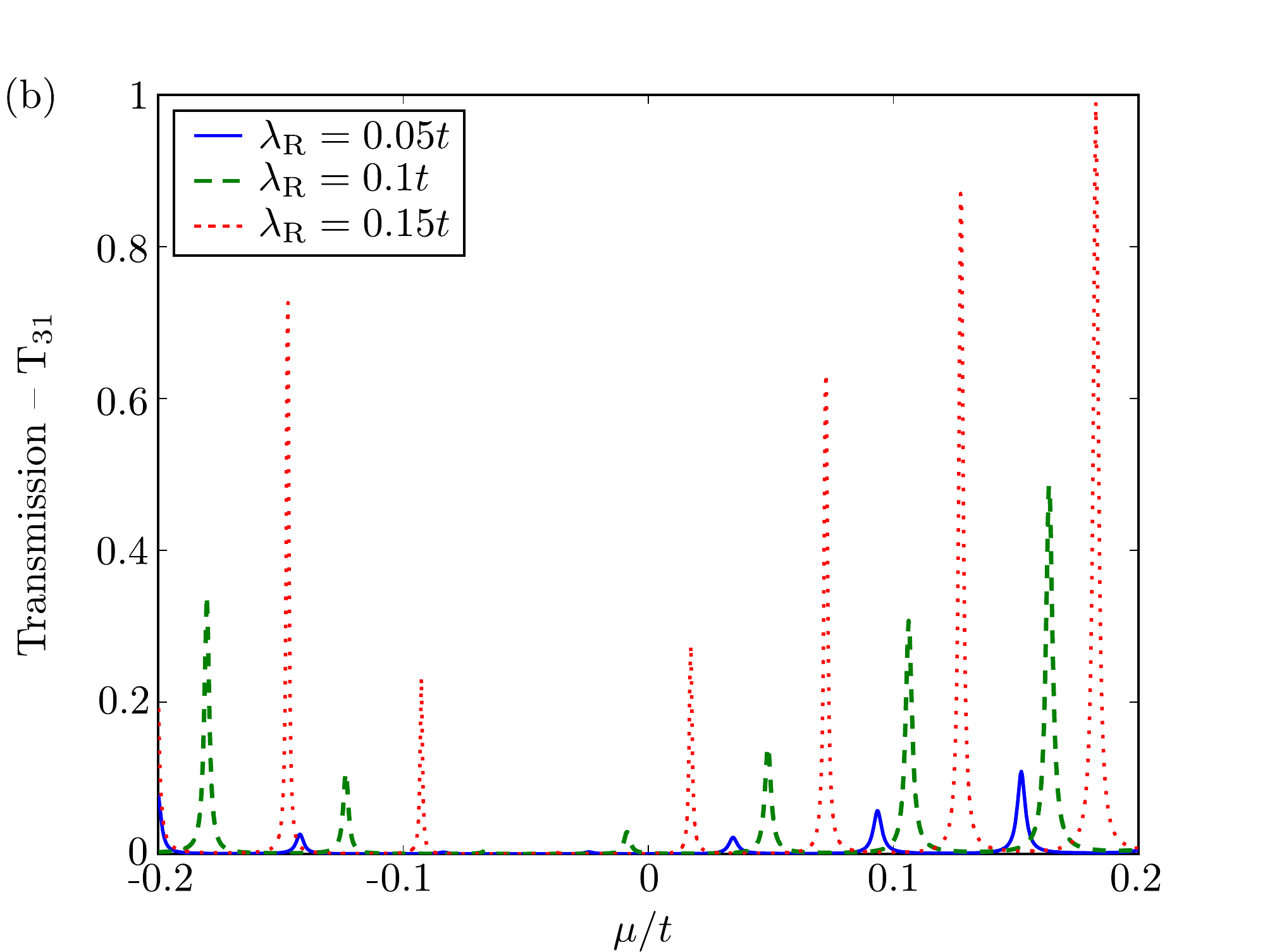}
\includegraphics[width=\columnwidth]{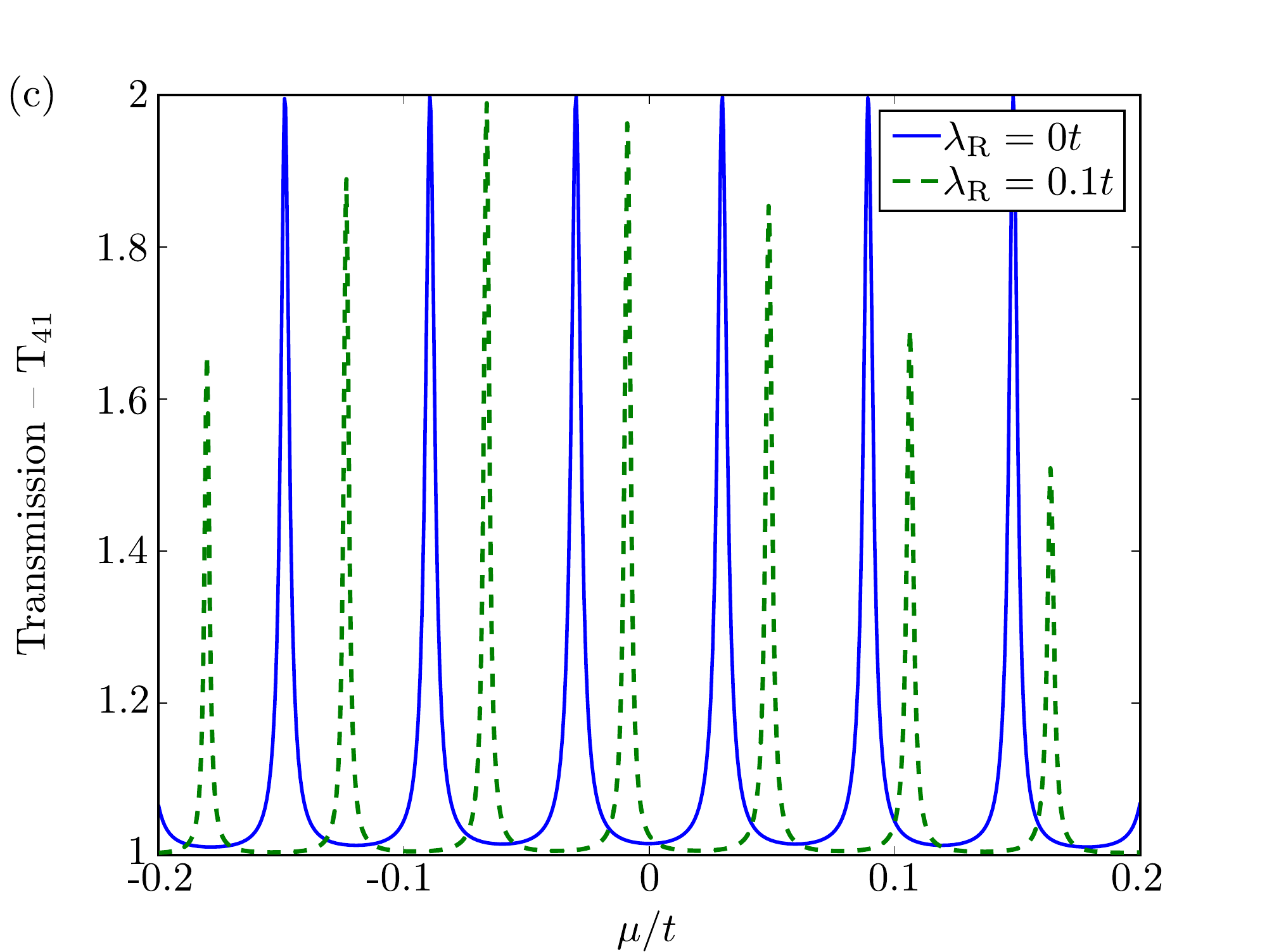}
\caption{Transmission coefficient from lead 1 to lead 2 on panel (a), to lead 3 on panel (b) and to lead 4 on panel (c), for several values of RSOC. The lattice setup is the same as in Fig.~\ref{fig:wavefunction}. The model parameters are set to: $t=1$, $\lambda_{\rm SO}=0.2t$.}
\label{fig:transmission}
\end{figure}

The most important effect of RSOC is to enable spin flips at the tunnel contacts. The amplitude for spin-flip processes can be quantified by calculating the transmission probability from lead $1$ to lead $i$, $T_{i1}$, shown in Fig.~\ref{fig:transmission} as a function of the chemical potential. On resonance with an antidot energy level, $T_{21}$ drops to zero, as shown in Fig.~\ref{fig:transmission}(a), in agreement with the analytic result in Eq.~(\ref{eq:S}) at $\phi=0$, showing that the injected electrons are fully transmitted to the opposite edge.
The transmission $T_{31}$ between leads 1 and 3, shown in Fig.~\ref{fig:transmission}(b), is only nonzero if there is spin-flip tunneling. Hence, in the absence of RSOC, all electrons are transmitted to lead 4 at resonance, as shown by the blue peaks in Fig.~\ref{fig:transmission}(c).
Moreover, the transmissions in Fig.~\ref{fig:transmission} are symmetric with respect to energy only in the absence of RSOC. In contrast, in the presence of RSOC, particle-hole symmetry is broken so that in general $T_{ij}(\mu)\neq T_{ij}(-\mu)$.

\begin{figure*}[t!]
\centering
\includegraphics[width=\textwidth]{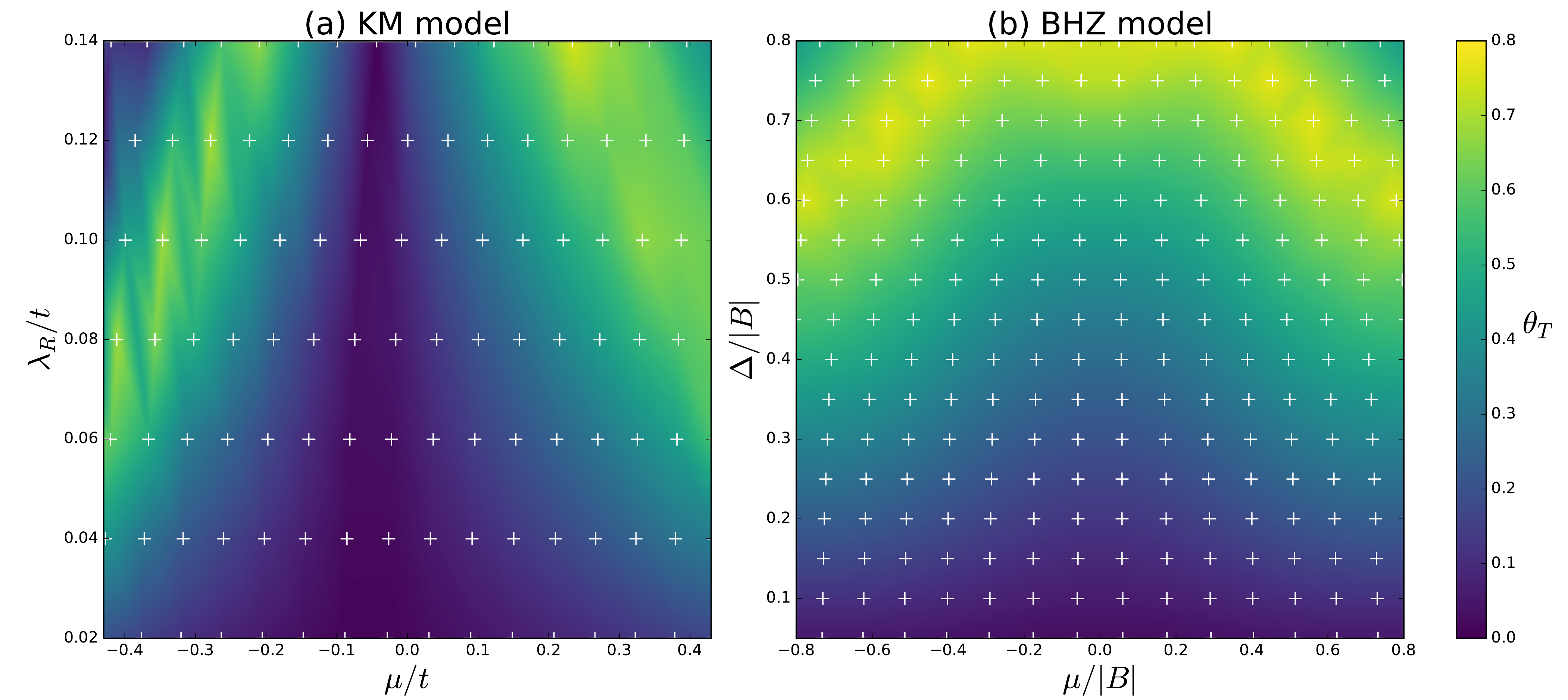}
\caption{(Color online) $\theta_T$ at the resonance energies for several values of (a) RSOC in the KM model and (b) bulk inversion asymmetry in the BHZ model. The white crosses indicate the points of transport resonance which were used to evaluate $\theta_T$. The model parameters are set to: $t=1$, $\lambda_{\rm SO}=0.2t$ for KM model, and $A=3$, $B=-1$, $M=-2$ and $C=D=0$ for BHZ model.\citep{bernevig-06s1757}}
\label{fig:theta_dependencies}
\end{figure*}

The peaks in the transmission probabilities all have Lo\-rentz\-ian shape around the resonance energies, but their widths change with energy. Using the numerical results for $T_{31}(\mu)$ or $T_{41}(\mu)$, we can calculate the values of $\theta_T$, shown in Fig.\ref{fig:theta_dependencies}(a), and $\gamma_T$ as functions of chemical potential. They turn out to vary slowly on the scale of the antidot level spacing. Hence, we are able to extract from the numerical simulations the dependence of the parameters of the analytic models on the tight-binding parameters as well as on chemical potential. Moreover, the Fermi velocity can be extracted from the band structure in the leads. Hence, we have access to all the quantities entering our analytic model via the numerical simulation.

We also performed tight-binding calculations based on the square-lattice discretization of the Bernevig-Hughes-Zhang model \citep{bernevig-06s1757} with added bulk inversion asymmetry,\cite{koenig-08jpsj031007} parametrized by $\Delta$, whose effect is similar to $\lambda_R$ in the KM model. The results do not differ qualitatively from the ones presented here for the KM model. The main difference consists of the preserved particle-hole symmetry as shown in Fig.~\ref{fig:theta_dependencies}(b).

\subsection{Non-local resistance}

\begin{figure}[t!]
\centering
 \includegraphics[width=\columnwidth]{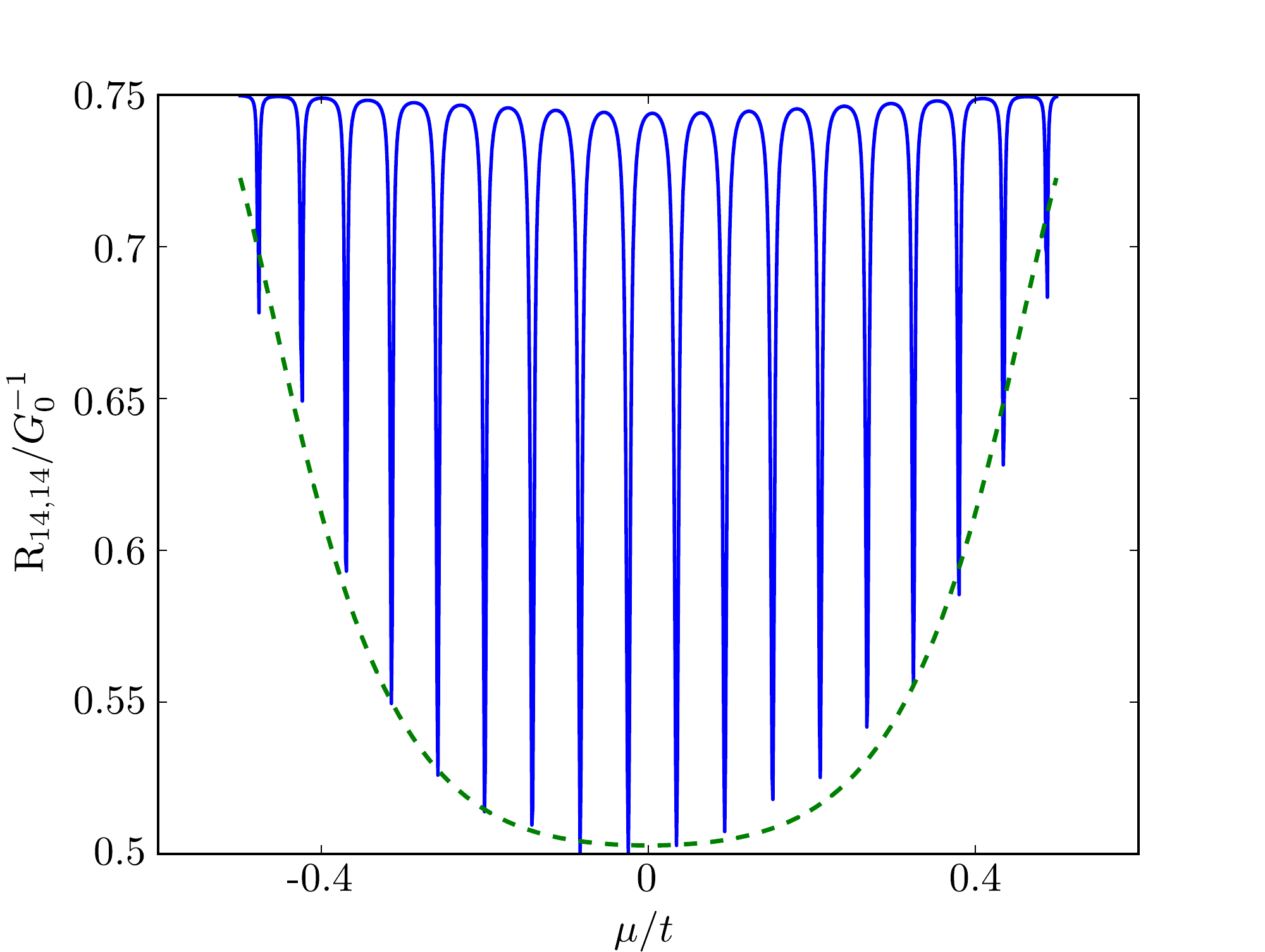}
\caption{Non-local resistance $R_{14,14}$ in unit of $G_0^{-1}$. The lattice setup is the same as in Fig.~\ref{fig:wavefunction}. The model parameters are set to: $t=1$, $\lambda_{\rm SO}=0.2t$ and $\lambda_{\rm R}=0.05t$. The continuous (blue) line is the computed non-local resistance for the lattice. The (green) dashed line is the fit of our model assuming that $\theta_T(\mu)\approx \alpha \mu^2 +\beta$.}
\label{fig:nonlocal_resistance}
\end{figure}

To compare the analytical predictions with the numerical simulations, we need to take into account the additional ballistic channels connecting contacts $1$ and $4$ and contacts $2$ and $3$ in Fig.~\ref{fig:setup} via the sample edges. This corresponds to replacing $T_{14}\to T_{14}+1$ and $T_{23}\to T_{23}+1$ (analogously for $T_{41}$ and $T_{32}$) obtained from Eq.~\eqref{eq:S}, while all other coefficients remain invariant.
The nonlocal multi-terminal resistance is then computed by means of the Landauer-B\"{u}ttiker formula~\eqref{eq:LB} in the linear-response regime. For instance, the relation between the current flowing between contacts $1$ and $4$ and the voltage developed at these same contacts is given by $R_{14,14}=(V_1-V_4)/I_1\vert_{I_4=-I_1,I_2=I_3=0}$. A numerical result is shown in Fig.~\ref{fig:nonlocal_resistance}.
In the homogeneous case $\theta_{\mathrm{U}k}=\theta_{\mathrm{L}k}$ and at resonance ($\sin\phi=0$) one finds
\begin{equation}
R_{14,14}=\left[\frac{1}{\cos(2\theta_T)+3}+\frac{1}{4}\right]G_0^{-1}.\label{non-local-resistance}
\end{equation}
In the absence of spin-flip tunneling ($\theta_T=0$), we find $R_{14,14}=(2G_0)^{-1}$. On the other hand, if only spin-flip processes are allowed ($\theta_T=\pi/2$), then $R_{14,14}=3/(4G_0)$.

As shown in Fig.~\ref{fig:nonlocal_resistance}, at resonance the nonlocal resistance reaches its minimum. Away from resonance, it tends towards $3/(4G_0)$. Around $\mu=0$, the nonlocal resistance deviates slightly from its quantized value due to the finite length of the tunneling region in the numerical simulation, which has the tendency to open a small spectral gap.\cite{sternativo14prb89_35415,chu09} We can fit Eq.~\eqref{non-local-resistance} to the envelope of the resonant peaks (green dashed line in Fig.~\ref{fig:nonlocal_resistance}) and thus determine the leading behavior $\theta_T(\mu) - \theta_T(0) \sim \mu^2$ for small $\mu$.

Let us briefly conclude the discussion of the non-interacting transport properties. In the case of an inhomogeneous RSOC, the transport properties depend explicitly on the spin texture. In the homogeneous case, the spin texture still appears implicitly in $\theta_T$. With the help of the numerical simulation, we are able to extract the parameters of the analytic model as a function of the bulk parameters and the chemical potential.

\section{Interacting antidot}\label{sec:int}

In confined low-dimensional systems, electron interactions are known to play an important role. Therefore, to complete the study of the transport properties through the QSH antidot, we need to investigate how interactions affect the transport mechanisms.
Computing the transport properties in the presence of electron interactions is generally a difficult task that cannot be solved exactly for arbitrary tunneling strength. Therefore, we focus on the lowest-order contributions to the tunneling current through the antidot, which are (A) sequential tunneling and (B) cotunneling.\cite{bruus_book}

\subsection{Sequential tunneling}
If the dwell time $\tau$ of electrons on the antidot is large such that $1/\tau\ll \{eV_i,k_BT\}$, the dominant transport processes are single-electron transfers between the edges and the antidot.
The transport properties can then be evaluated within first order perturbation theory in $\vert\gamma_T\vert^2$.
We assume that the antidot contains $N$ electrons in the ground state, and we assume $N$ to be even without loss of generality. The initial state of the leads contains one electron at a certain momentum $k$ in one of the leads. The initial state of the full system is thus a direct product of the initial state in the leads and in the antidot:
\begin{equation}
|i^{s\alpha}(N, k)\rangle=|N\rangle\otimes c^\dag_{s\alpha k}\vert \text{vac} \rangle.
\end{equation}
We compute the transition rate for adding another electron on the antidot. The final state due to tunneling of one electron from the edge $s$ with initial momentum $k$ and chirality $\alpha$ to the antidot, with final chirality $\alpha^{\prime}$ and angular momentum $j$, reads
\begin{equation}
|f^{s\alpha\alpha^{\prime}}(N+1,k,j)\rangle=d^{\dagger}_{\alpha^{\prime}j}c_{s\alpha k}\vert i^{s\alpha}(N,k)\rangle
.\end{equation}
According to Fermi's golden rule, the rate for transitions from the initial state to the final state is\cite{bruus_book}
\begin{equation}
\Gamma_{N+1, N}^{s\alpha\alpha^\prime}(k,j)=2\pi |\langle f|\H_{\mathrm{d},s}|i\rangle|^2F_i\delta(E_{f}-E_{i})
,\end{equation}
where $E_f - E_{i}= E(N+1) - E(N) +\varepsilon_{\mathrm{d}\alpha^{\prime}}(j)-\varepsilon_{s\alpha}(k)$ is the energy difference between final and initial states and $F_i$ is a Fermi function denoting the probability of finding the system in the initial state $|i\rangle$. $\varepsilon_{\mathrm{d}\alpha^{\prime}}(j)$ is the eigenenergy of the antidot with angular momentum $j$ and chirality $\alpha^\prime$, and $\varepsilon_{s\alpha}(k)$ is the eigenenergy of the $s$ edge with momentum $k$ and chirality $\alpha$. The total transition rate is then obtained by summing over all possible initial and final states,
\begin{equation}\label{eq:rate_seq}
\Gamma_{N+1, N}^{s\alpha\alpha^\prime}=\sum_{j,k} \Gamma_{N+1, N}^{s\alpha\alpha^\prime}(k,j)
.\end{equation}
In the sequential tunneling regime, the tunneling current is evaluated using a rate-equation approach.
In the dc limit, and by considering only two antidot states with either $N$ or $N+1$ electrons, which is valid as long as the charging energy is large enough to forbid other occupation numbers, one has
\begin{equation}
\Gamma_{{N+1}, {N}}P(N)=\Gamma_{{N}, {N+1}}P({N+1}) \label{recursive_prob_eq}
\end{equation}
with $\Gamma_{N+1,N}=\sum_{s,\alpha,\alpha^{\prime}}\Gamma_{{N+1}, {N}}^{s\alpha\alpha^{\prime}}$ and similarly for $\Gamma_{N,N+1}$. Combined with the conservation of probability constraint, $P(N)+2P(N+1)=1$, it is then possible to compute the occupation probabilities in terms of the transition rates. Transitions between the state with $N$ electrons and the one with $N+1$ electrons in the antidot are enabled close to the resonance condition $\Delta E(N+1)=E(N+1)-E(N) = 0$. Therefore, transitions between $N$ and $N+1$ states are allowed for $n_g \approx N+\frac{1}{2}$, where $n_g=eV_g/E_c$ is determined by the gate voltage and the charging energy.
The recursive equation \eqref{recursive_prob_eq} and the probability conservation yield the expression of the probabilities, which are necessary to compute the total current,
\begin{equation}
I=-e\left[P(N)\Gamma_{N+1,N}^{{\mathrm U}}-P(N+1)\Gamma_{N,N+1}^{{\mathrm U}}\right]. \label{seq_current}
\end{equation}
The expression of the current is still complicated and depends on how many levels can be reached in the bias window. If the bias window is sufficiently small, tunneling is only possible via one energy level  $\varepsilon_{{\mathrm d}}(j)$ situated near the Fermi energy $\mu$. In this case,
\begin{equation}
I=-e\frac{2\Gamma_T}{\pi R}\frac{ n_F(\varepsilon_{{\mathrm d}}(j)-\mu_{\mathrm U })-n_F(\varepsilon_{{\mathrm d}}(j)-\mu_{\mathrm L})}{2+\left[n_F(\varepsilon_{{\mathrm d}}(j)-\mu_{\mathrm U })+n_F(\varepsilon_{{\mathrm d}}(j)-\mu_{\mathrm L })\right]},
\end{equation}
where $\Gamma_T=|\gamma_T|^2/(2v_F) $. We choose the chemical potentials as $\mu_{\mathrm U}=\mu+eV/2$ and $\mu_{\mathrm L}=\mu-eV/2$, where $V$ is the bias voltage between upper and lower edges. At $T=0$ and finite voltage, we get
\begin{equation}
I_{T=0}=-\frac{2e\Gamma_T}{3\pi R},
\end{equation}
for $\varepsilon_{{\mathrm d}}(j)\in[\mu-eV/2,\mu+eV/2]$. On the other hand, if the temperature is finite and $eV \ll k_B T$, the current to lowest order in the applied voltage becomes
\begin{equation}
I_{T\neq 0}=-\frac{1}{4k_BT}\frac{\Gamma_T}{\pi R}\frac{e^2V}{1+n_F(\tilde{\varepsilon}_{{\mathrm d}})}\frac{1}{\cosh^2(\tilde{\varepsilon}_{{\mathrm d}}/(2k_BT))},
\end{equation}
where $\tilde{\varepsilon}_{{\mathrm d}}=\varepsilon_{{\mathrm d}}(j)-\mu$. As is well known in the sequential tunneling regime, the limits $eV \to 0$ and $k_B T \to 0$ do not commute.\cite{bruus_book}

One of the central assumptions of this rate equation approach is that the electrons on the dot relax to the ground state between tunneling events, i.e., there is a separation of time scales between the fast relaxation and the slow tunneling. However, in our case, either the initial state $|N\rangle$ or the final state $|N+1\rangle$ is twofold-degenerate due to time-reversal symmetry. Since the rate-equation approach does not properly account for the fact that the chirality of the electrons on the antidot is conserved, it is not possible to calculate chirality-resolved currents within this approach. Hence, we only presented results for the total current $I$. However, since the total current does not contain information about the spin texture, we continue by exploring higher-order coherent processes.

\subsection{Cotunneling}

\begin{figure}[t!]
\centering
\includegraphics[scale=0.5]{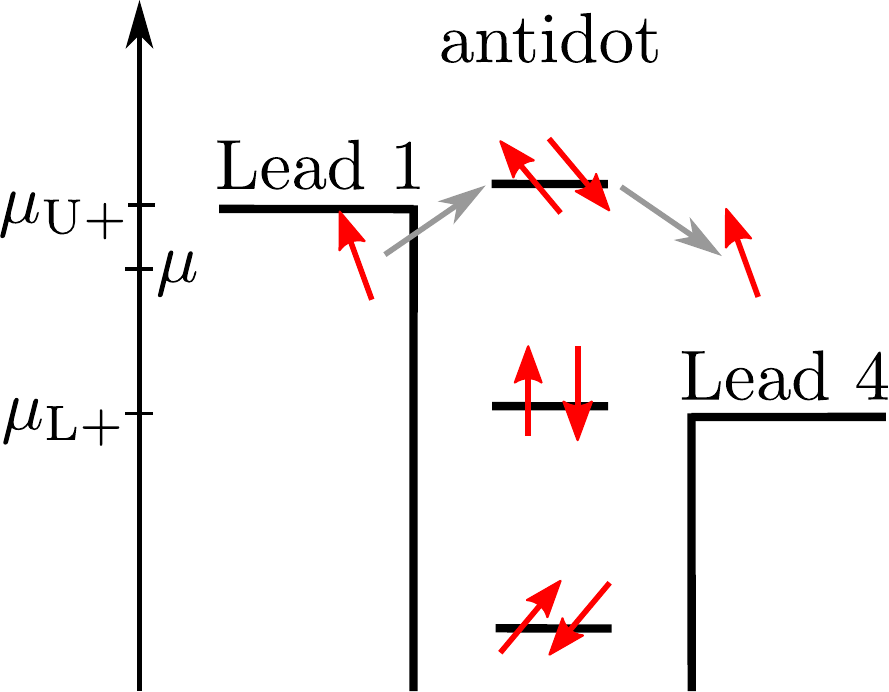}
\caption{Sketch of a possible cotunneling process between lead 1 and lead 4. The energy level on the antidot is fully occupied. One of the two electrons escape the antidot towards lead 4, creating a virtual state with an additional hole. The final state is reached when the electron in lead 1 tunnels to the antidot.}
\label{fig:cotunneling}
\end{figure}

If sequential tunneling is inhibited due to energy conservation, transport between the upper and lower edge is still possible via cotunneling, where electrons tunnel between the upper and lower edge via virtual states on the antidot. A sketch of a possible process is shown in Fig.~\ref{fig:cotunneling}.
In the following we consider elastic cotunneling which, due to the discrete nature of the mesoscopic antidot, becomes relevant for transport.
The cotunneling rate from the initial state $\vert i^{s\alpha}(N,k)\rangle$, which contains $N$ electrons on the antidot and a single electron with momentum $k$ and chirality $\alpha$ on the edge $s$, to the final one $\vert f^{s'\alpha'}(N,k')\rangle$, which is defined analogously, reads
\begin{equation}
\begin{split}
&\Gamma_{i\to f}=2\pi\delta(E_{f}-E_{i})F_i\\
&\times|\langle f|\H_{\mathrm{d,L}}\frac{1}{E_{i}-\H_{0}}\H_{\mathrm{d,U}}+\H_{\mathrm{d,U}}\frac{1}{E_{i}-\H_{0}}\H_{\mathrm{d,L}}|i\rangle|^2,\label{cotunneling_total}
\end{split}
\end{equation}
where $\H_{0}=\H_{ {\mathrm U}}+\H_{{\mathrm L}}+\H_{ {\mathrm d}}$, and $E_{i,f}$ are the energies of the initial and final state. $F_i$ is again the Fermi distribution specifying the probability of finding the system in the initial state $|i^{s\alpha}(N,k)\rangle$. Note that the antidot contains $N$ electrons in the ground state in both initial and final states.

The total chirality-resolved cotunneling rates $\Gamma^{\alpha\alpha '}_{\mathrm{U}\to\mathrm{L}}$ and $\Gamma^{\alpha\alpha '}_{\mathrm{L}\to\mathrm{U}}$ are then obtained by summing over all the possible initial- and final-state momenta, and over all angular momenta and chiralities in the intermediate state.
From the tunneling rates it is then possible to compute the tunneling current, defined as flowing from the upper to the lower edge, as
\begin{equation}\label{eq:cot_curr}
I^{\alpha \alpha^\prime}=(-e)\left[\Gamma^{\alpha\alpha^\prime}_{\mathrm{U}\to\mathrm{L}}-\Gamma^{\alpha ^\prime\alpha}_{\mathrm{L}\to\mathrm{U}}\right].\end{equation}
To connect to the setup shown in Fig.~\ref{fig:setup}, $I^{++}$ is the current flowing from lead 1 to lead 4, $I^{+-}$ is the current flowing from lead 1 to lead 3, $I^{-+}$ is the current flowing from lead 2 to lead 4, and $I^{--}$ the current flowing from lead 2 to lead 3. The results for these currents depend strongly on the parity of the antidot occupation.

In the case of an even number $N$ of electrons on the antidot, i.e., if all levels up to the chemical potential are doubly occupied, we find
\begin{equation}
\begin{split}
I_{N\ \rm even}^{\alpha\alpha^\prime}=&-\frac{e\Gamma_T^2 }{2\pi^3R^2}\sum_{j}\int d\varepsilon [n_F(\varepsilon-\mu_{{\mathrm U}\alpha})-n_F(\varepsilon-\mu_{{\mathrm L}\alpha^\prime})]\\
&\times\frac{ \cos^2(2\theta_T)\delta_{\alpha^\prime\alpha}+\sin^2(2\theta_T)\delta_{\bar{\alpha}^\prime\alpha} }{\left [\varepsilon-\varepsilon_{{\mathrm d}}(j)-\Delta E\right ]^2}
,\end{split}
\end{equation}
where $\varepsilon_{{\mathrm d}}(j)+\Delta E$ is the energy of the intermediate state. Depending on the gate voltage, $\Delta E= \Delta E (N)$ if an additional hole occupies the $j$-th level of the dot or $\Delta E= \Delta E (N+1)$ for an additional electron.
At low bias it is possible to expand the Fermi functions around $\mu$ and, by assuming that only the $j_0$th antidot energy level (the closest from the chemical potential) is contributing to the cotunneling current and $\Delta E+\varepsilon_{{\mathrm d}}(j_0)\gg \mu_{\mathrm{U/L}\alpha}$, one obtains
\begin{equation}
\begin{split}
I_{N\ \rm even}^{\alpha\alpha^\prime}\approx&-\frac{e^2(V_\alpha-V_{\alpha^\prime})\Gamma_T^2 }{2\pi^3R^2}
\frac{ \cos^2(2\theta_T)\delta_{\alpha^\prime\alpha}+\sin^2(2\theta_T)\delta_{\bar{\alpha}^\prime\alpha} }{\left [\mu-\varepsilon_{{\mathrm d}}(j_0)-\Delta E\right ]^2}
\end{split}
\end{equation}
with the parameter $\theta_T$ evaluated at the chemical potential $\mu$. We recover here an implicit dependence on the spin texture through $\theta_T$. The Coulomb repulsion will just shift the energy in the denominator, depending on the value of the gate voltage. Moreover, it is possible to invert the expression for the current in order to extract the value of $\theta_T$ as a function of the current, the voltage, and the charging energy. This makes it possible in principle to compare the tunneling processes in the noninteracting and in the interacting limits.

The case in which the antidot level hosts an odd number of electrons, i.e., if the energy level at the chemical potential has only one electron, is quite different.
Due to the degeneracy of the antidot energy level, the initial and final states should include the initial (final) chirality of the antidot $\beta^{(\prime)}$, becoming $|i^{s\alpha\beta}(N,k)\rangle$ and $\vert f^{s'\alpha'\beta^\prime}(N,k')\rangle$.\cite{lehmann06} By applying Eq.~\eqref{cotunneling_total}, we modify the current expression in \eqref{eq:cot_curr} to
\begin{equation}\label{eq:mod_cot_curr}
I^{\alpha \alpha^\prime}=(-e)\sum_{\beta,\beta^\prime}\left[\Gamma^{\alpha\alpha^\prime\beta\beta^\prime}_{\mathrm{U}\to\mathrm{L}}-\Gamma^{\alpha ^\prime\alpha\beta\beta^\prime}_{\mathrm{L}\to\mathrm{U}}\right]p_\beta.\end{equation}
The rates are found in Appendix~\ref{appendix:rates}. The probabilities $p_\beta$ of the highest level of the antidot being occupied by an electron with chirality $\beta$ are determined by the conservation of probabilities $p_++p_-=1$ and by the rate equation,
\begin{equation}
\frac{dp_\beta}{dt}=-\Gamma^{\beta\bar{\beta}}p_\beta+\Gamma^{\bar{\beta}\beta}p_{\bar{\beta}}=0,
\end{equation}
where $\Gamma^{\beta\bar{\beta}}=\sum_{\alpha\alpha^\prime}(\Gamma^{\alpha\alpha^\prime\beta\bar{\beta}}_{\mathrm{U}\to\mathrm{L}}+\Gamma^{\alpha\alpha^\prime\beta\bar{\beta}}_{\mathrm{L}\to\mathrm{U}})$. By evaluating the rates at the chosen chemical potentials, we are able to compute the chirality-resolved currents. As an example, we set a difference of potential between the upper edge and the lower edge, such that $\mu_{\mathrm{U}+}=\mu_{\mathrm{U}-}=\mu+\frac{eV}{2}$ and $\mu_{\mathrm{L}+}=\mu_{\mathrm{L}-}=\mu-\frac{eV}{2}$, leading to
\begin{equation}
\begin{split}
I_{N\ \rm odd}^{\alpha\alpha^\prime}\approx-\frac{2\Gamma_T^2G_0V}{v_F^2}&\left [\frac{1}{\xi^2(N)}+\frac{1}{\xi^2(N+1)}\right .\\
&\left .+\frac{\alpha\alpha^\prime\cos\left (4\theta_{k_F}-4\theta_{j_0}\right)-1}{\xi(N)\xi(N+1)}\right ]
,\end{split}
\end{equation}
where $\xi(N)=2\pi R[\varepsilon_{{\mathrm d}}(j_0)+\Delta E(N)-\mu]/v_F$. In this result, we obtain an explicit dependence of the current on the antidot and edge state spin textures, $\theta_{j_0}$ and $\theta_{k_F}$, even in the case of position-independent RSOC. Hence, a measurement of the cotunneling current allows one to measure the difference between the external edge spin rotation $\theta_{k_F}$ and the antidot spin rotation $\theta_j$. Since the system occupies a virtual intermediate state, $\theta_{k_F}$ will generally differ from $\theta_{j_0}$. In particular, the difference of currents between the two lower terminals leads to
\begin{equation}
I_{N\ \rm odd}^{+-}-I_{N\ \rm odd}^{++}\approx \left (\frac{2\Gamma_T}{v_F}\right )^2\frac{\cos\left (4\theta_{k_F}-4\theta_{j_0}\right)}{\xi(N)\xi(N+1)}G_0V
\end{equation}
This measurement would enable us to extract directly the information about the spin texture. 

\begin{figure}[t!]
\centering
\includegraphics[scale=0.47]{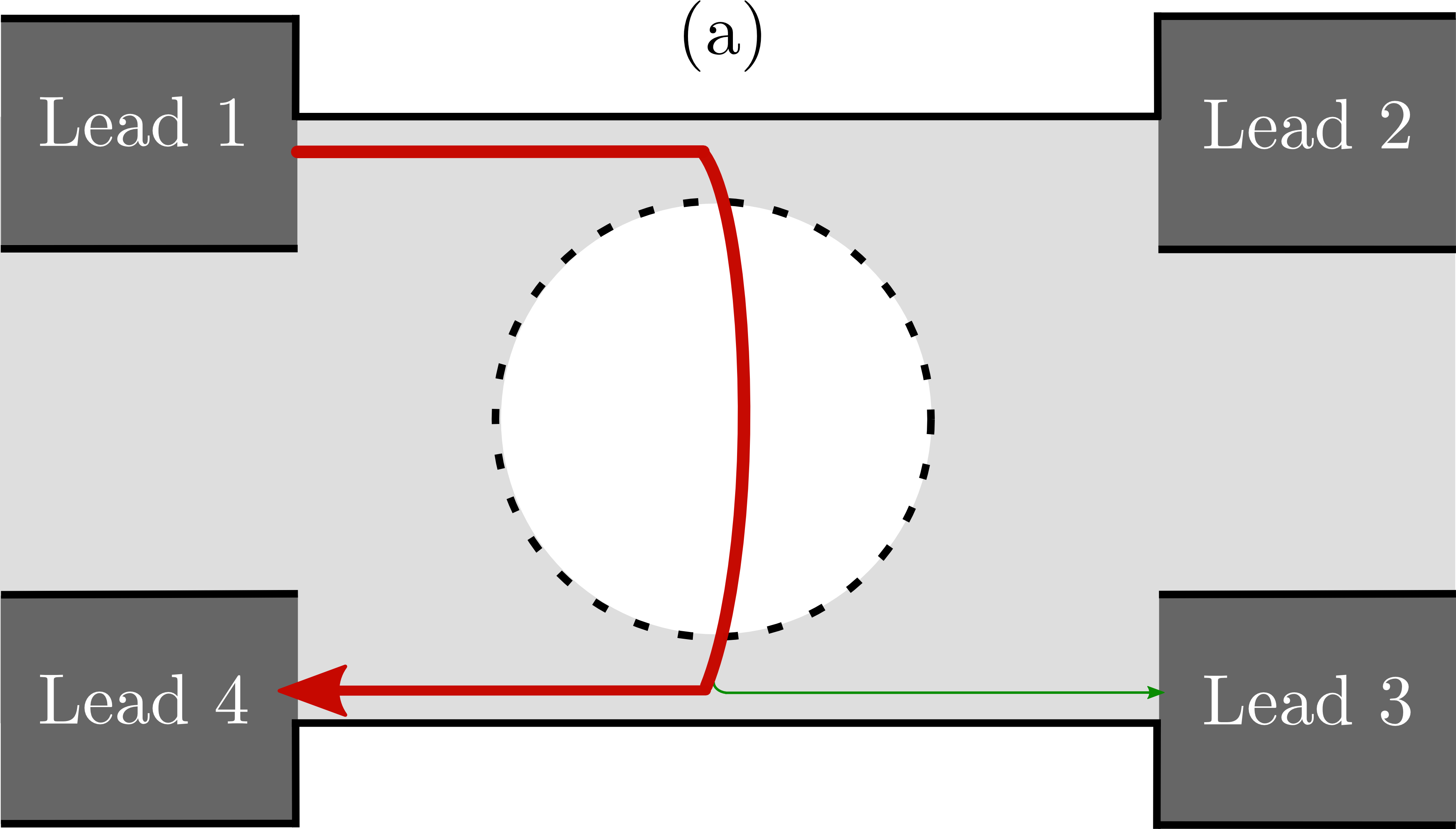}
\includegraphics[scale=0.47]{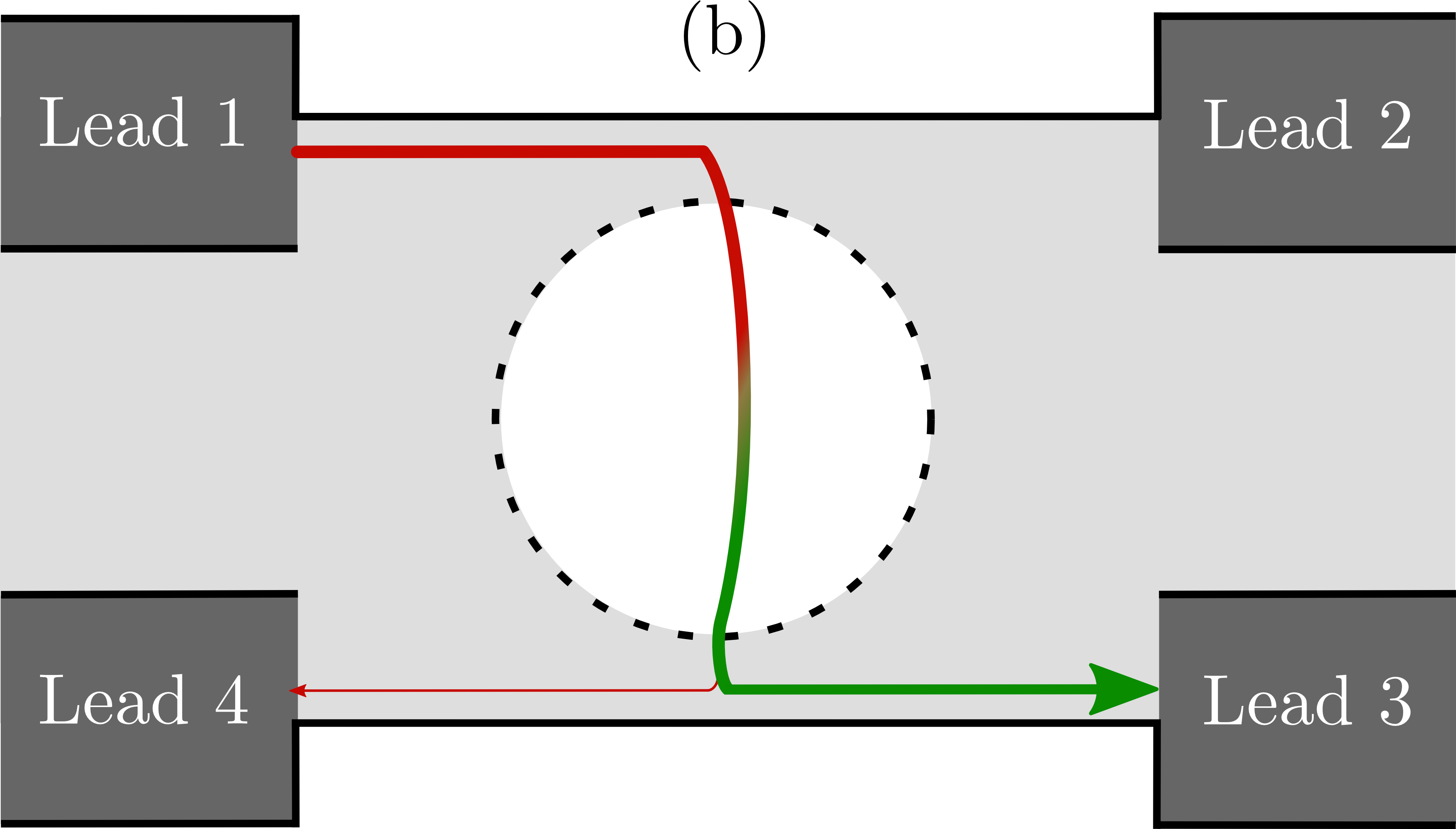}
\caption{Sketch of the two contributions to the tunneling current $I_{N\ \rm odd}^{++}$ and $I_{N\ \rm odd}^{+-}$ in the case only lead $1$ is biased, that is $\mu_{U+}=\mu +eV$ and $\mu_{U-}=\mu_{L+}=\mu_{L-}=\mu$.
Panel (a) corresponds to weak Rashba interactions, so that spin is almost conserved: therefore the contribution $I_{N\ \rm odd}^{+-}$ due to spin-flip is strongly suppressed. In the opposite scenario of strong Rashba interactions, spin-flip contributions can even become dominant compared to spin-preserving one, leading to the scenario depicted in panel (b).}
\label{fig:cotunneling_setup}
\end{figure}

Another possibility to probe the spin texture would be to apply a different voltage setting: only one lead is biased, such as for example $\mu_{U+}=\mu +eV$ and $\mu_{U-}=\mu_{L+}=\mu_{L-}=\mu$. Again, we can divide the tunneling current in two contributions $I_{N\ \rm odd}^{++}$ and $I_{N\ \rm odd}^{+-}$ that we compute using Eq.~\ref{eq:mod_cot_curr}, leading to
\begin{widetext}
\begin{eqnarray}\label{eq:iod+-}
I_{N\ \rm odd}^{+-}&\approx&-\frac{2\Gamma_T^2G_0V}{v_F^2}\left (\frac{1}{\xi(N+1)\xi(N)}\right )^2 \left \{\xi^2(N+1)+\xi^2(N)-\xi(N+1)\xi(N)\left [1+\cos\left (4\theta_{k_F}-4\theta_{j_0}\right )\right ]\right .\nonumber\\
&-&\left .[\xi(N+1)-\xi(N)]\cos\left (2\theta_{k_F}-2\theta_{j_0}+2\theta_T\right )\left [\xi(N+1)\cos\left (2\theta_{k_F}-2\theta_{j_0}-2\theta_T\right )-\xi(N)\cos\left (2\theta_{k_F}+2\theta_{j_0}+2\theta_T\right )\right ]\right \}
\end{eqnarray}
\begin{eqnarray}\label{eq:iod++}
I_{N\ \rm odd}^{++}&\approx&-\frac{2\Gamma_T^2G_0V}{v_F^2}\left (\frac{1}{\xi(N+1)\xi(N)}\right )^2 \left \{\xi^2(N+1)+\xi^2(N)-\xi(N+1)\xi(N)\left [1-\cos\left (4\theta_{k_F}-4\theta_{j_0}\right )\right ]\right .\nonumber\\
&+&\left .[\xi(N+1)-\xi(N)]\cos\left (2\theta_{k_F}-2\theta_{j_0}+2\theta_T\right )\left [\xi(N+1)\cos\left (2\theta_{k_F}-2\theta_{j_0}-2\theta_T\right )+\xi(N)\cos\left (2\theta_{k_F}-2\theta_{j_0}+2\theta_T\right )\right ]\right \}
.\end{eqnarray}
\end{widetext}
We observe this time a more sophisticated dependence on the different parameters.
However, from Eq.~\eqref{eq:iod+-} one can see that the current flowing to lead $3$ vanishes if $\theta_{k_F}=\theta_{j_0}=\theta_T=0$, that is, in the absence of RSOC; indeed, in this case spin is conserved and electrons, injected fully spin-up polarized from lead 1, can only flow to lead 4 preserving their spin, as schematically shown in Fig.~\ref{fig:cotunneling_setup}(a).
However, in the case of strong RSOC, spin-flip tunneling can become important, eventually dominating over the spin-preserving contribution. In this case, the current mostly flows from lead $1$ to lead $3$, as schematically shown in Fig.~\ref{fig:cotunneling_setup}(b).

To conclude the discussion of the electron transport in the interacting case, we observe in the cotunneling regime that the occupation of the antidot plays an important role. In the case of an even occupation, we recover the implicit dependence of the current on the spin texture. In contrast, for odd occupation, we find an explicit dependence on the spin texture. By tuning the chemical potentials, it is thus possible to detect the interplay between implicit ($\theta_T$) and explicit ($\theta_{k_F}$ and $\theta_j$) dependence of the spin texture.

\section{Conclusion}\label{sec:conclusions}

We have presented a detailed analysis of the electron transport between the edges of a two-dimensional topological insulator via an antidot. In particular, we investigated the effects of a nontrivial spin structure of the edge states and a charging energy due to Coulomb repulsion on the antidot.

We first presented a solution of the corresponding scattering problem in the absence of interactions. We showed that, on the one hand, spin-non-conservation modifies the spin texture of the edge states, but on the other hand, it also makes spin-flip tunneling between the edges and the antidot possible. We also performed numerical calculations based on tight-binding models, and we confirmed the predictions of the scattering approach. We found that the effects of spin-non-conservation become most important for inhomogeneous samples where the spin structures of different edge states may differ.

To include the effects of charging energy, we investigated the Coulomb blockade regime, where we presented a calculation of the sequential tunneling and cotunneling currents. Here, we showed in particular that, since the cotunneling occurs via an intermediate, virtual state on the antidot, it allows a spectroscopic measurement of the antidot and edge-state spin structure.

\begin{acknowledgments}
This work was supported through the German DFG priority program SPP 1666 on topological insulators. AR, GD, and TLS acknowledge support from the National Research Fund, Luxembourg (ATTRACT 7556175). SR thanks the DFG for financial support through SFB 1143.
\end{acknowledgments}

\appendix
\begin{widetext}
\section{Cotunneling rates for an odd number of electrons}\label{appendix:rates}
When we apply Eq.~\eqref{cotunneling_total}, with the additional degree of freedom due to the chirality, we are able to compute the 16 transition rates appearing in Eq.~\eqref{eq:mod_cot_curr}.
\begin{equation}
\begin{split}
\Gamma^{\alpha\alpha\beta\beta}_{\mathrm{U}\to\mathrm{L}}=\frac{\Gamma_T^2}{2\pi v_F^2}&\sum_j\int d\varepsilon n_F(\varepsilon-\mu_{U\alpha})[1-n_F(\varepsilon-\mu_{L\alpha})]
\left[\frac{\cos(2\theta_T)+\alpha\beta\cos(2\theta_k-2\theta_j)}{\xi(N)+\mu-\varepsilon}\right.
\left.-\frac{\cos(2\theta_T)-\alpha\beta\cos(2\theta_k-2\theta_j)}{\xi(N+1)+\mu-\varepsilon}\right]^2,
\end{split}
\end{equation}
\begin{equation}
\begin{split}
\Gamma^{\alpha\alpha\beta\bar{\beta}}_{\mathrm{U}\to\mathrm{L}}=\frac{\Gamma_T^2}{2\pi v_F^2}&\sum_j\int d\varepsilon n_F(\varepsilon-\mu_{U\alpha})[1-n_F(\varepsilon-\mu_{L\alpha})]
\left[\frac{\sin(2\theta_T)+\alpha\beta\sin(2\theta_k-2\theta_j)}{\xi(N)+\mu-\varepsilon}
\right.
\left.-\frac{\sin(2\theta_T)+\alpha\beta\sin(2\theta_k-2\theta_j)}{\xi(N+1)+\mu-\varepsilon}\right]^2,
\end{split}
\end{equation}
\begin{equation}
\begin{split}
\Gamma^{\alpha\bar{\alpha}\beta\beta}_{\mathrm{U}\to\mathrm{L}}=\frac{\Gamma_T^2}{2\pi v_F^2}&\sum_j\int d\varepsilon n_F(\varepsilon-\mu_{U\alpha})[1-n_F(\varepsilon-\mu_{L\bar{\alpha}})]
\left[\frac{\sin(2\theta_T)-\alpha\beta\sin(2\theta_k-2\theta_j)}{\xi(N)+\mu-\varepsilon}\right.
\left.-\frac{\sin(2\theta_T)+\alpha\beta\sin(2\theta_k-2\theta_j)}{\xi(N+1)+\mu-\varepsilon}\right]^2,
\end{split}
\end{equation}
\begin{equation}
\begin{split}
\Gamma^{\alpha\bar{\alpha}\beta\bar{\beta}}_{\mathrm{U}\to\mathrm{L}}=\frac{\Gamma_T^2}{2\pi v_F^2}&\sum_j\int d\varepsilon n_F(\varepsilon-\mu_{U\alpha})[1-n_F(\varepsilon-\mu_{L\bar{\alpha}})]
\left[\frac{\cos(2\theta_T)-\alpha\beta\cos(2\theta_k-2\theta_j)}{\xi(N)+\mu-\varepsilon}\right.
\left.-\frac{\cos(2\theta_T)-\alpha\beta\cos(2\theta_k-2\theta_j)}{\xi(N+1)+\mu-\varepsilon}\right]^2.
\end{split}
\end{equation}

The remaining rates for the transition $\Gamma^{\alpha^\prime\alpha\beta\beta^\prime}_{\mathrm{L}\to\mathrm{U}}$ are obtained by detailed balance.
\end{widetext}

\end{document}